\documentclass[11pt]{article} 
\usepackage[margin=1in]{geometry}
\usepackage[T1,T2A]{fontenc}
\usepackage[cmex10]{amsmath}
\usepackage{amsthm}
\usepackage{algorithmicx, algpseudocode}
\usepackage[linesnumbered,ruled,vlined]{algorithm2e}
\newtheorem{mydef}{Definition}

\SetKwInput{KwData}{Input}
\SetKwInput{KwResult}{Output}

\usepackage{comment}
\usepackage{todonotes}
\usepackage{color,colortbl}
\usepackage[misc]{ifsym}
\usepackage{graphicx}
\graphicspath{ {./images/} }
\usepackage{subcaption}

\providecommand{\keywords}[1]
{
  \small	
  \textbf{\textit{Keywords---}} #1
}

\theoremstyle{definition}

\theoremstyle{remark}

\title{Threat from being Social: Vulnerability Analysis of Social Network Coupled Smart Grid}

\author{Tianyi Pan \thanks{T. Pan, L. N. Nguyen, and My T. Thai are with the Department
of Computer and Information Science and Engineering, University of Florida, Gainesville,
FL, 32611, USA. Email: \{tianyi,nglan,mythai\}@cise.ufl.edu}
\thanks{T. Pan and S. Mishra contributed equally to the paper}
\and Subhankar Mishra \thanks{S. Mishra is with the National Institute of Science Education and Research, Bhubaneswar, India.  Email: smishra@niser.ac.in} \footnotemark[2]
\and  Lan N. Nguyen \footnotemark[1]
\and Gunhee Lee \thanks{G Lee and J Kang are with National Security Research Institute, Daejeon, Republic of Korea. Email: \{icezzoco, jmkang\}@nsr.re.kr}
\and Jungmin Kang  \footnotemark[4]
\and Jungtaek Seo \thanks{Jungtaek Seo is with the Department of Information Security Engineering, Soonchunhyang University, Republic of Korea. Email: seojt@sch.ac.kr}   
\and My T. Thai \footnotemark[1]
}

\date{}
\begin{document}
\maketitle

\abstract{
    Social Networks (SNs) have been gradually applied by utility companies as an addition to smart grid and are proved to be helpful in smoothing load curves and reducing energy usage. However, SNs also bring in new threats to smart grid: misinformation in SNs may cause smart grid users to alter their demand, resulting in transmission line overloading and in turn leading to catastrophic impact to the grid. In this paper, we discuss the interdependency in the social network coupled smart grid and focus on its vulnerability. That is, how much can the smart grid be damaged when misinformation related to it diffuses in SNs? To analytically study the problem, we propose the Misinformation Attack Problem in Social-Smart Grid (MAPSS) that identifies the top critical nodes in the SN, such that the smart grid can be greatly damaged when misinformation propagates from those nodes. This problem is challenging as we have to incorporate the complexity of the two networks concurrently. Nevertheless, we propose a technique that can explicitly take into account information diffusion in SN, power flow balance and cascading failure in smart grid integratedly when evaluating node criticality, based on which we propose various strategies in selecting the most critical nodes. Also, we introduce controlled load shedding as a protection strategy to reduce the impact of cascading failure. The effectiveness of our algorithms are demonstrated by experiments on IEEE bus test cases as well as the Pegase data set. 
}
\\ \bigskip

\section{Introduction}
A smart grid is the next generation of power grid based on information technology and real-time data processing which allow the implementation of strategies to control and optimize the electric network.
The recent literature has not only focused on addressing the technological hardware and software features of the smart grid, but also on the social dimension of the grid \cite{Goulden2014,Stem2014}. In the past few years, media has been coining the idea of connecting the online social networks with smart grid and exploring it as combined research topic \cite{Boslet2010, Chima2011
	, Fang2013}. Social network and smart grid have been studied much individually; however, there exists limited work in the linked smart grid and social network. A number of recent papers propose frameworks or approaches that interconnect smart meters (or smart homes) as SNs for energy management and sharing \cite{Ciuciu2012,Stein2012}. In addition, several frameworks or simulation models for demand side management and value-added web services with social networking aspects have been developed \cite{Chat2013,Haan2011,Lei2012}. Several others \cite{Skopik2014,Worm2013} have used simulation models to demonstrate the feasibility of social coordination in supply and demand. Our research work focuses on the impact of social network on the smart grid, when they are linked together.

The social computing will integrate and enhance many digital systems over the next decade and the smart grid is no exception \cite{Woody2012,McDonald2011}. Smart grid efficiency depends on utility customers having knowledge about demand response programs and being actively engaged in energy management. And this is exactly where social network comes into the picture and can really have an impact. Social computing can also expand the adoption and adaptation of smart grid technologies through the peer to peer communication in local communities via social network. It also could change large scale behavior through crowd-shifting basing on the theory ``people decide how to behave based on what they see others doing, especially if those others seem similar to themselves" \cite{LaMonica2010,Raafat2009}. Through the social network, the utility companies can learn about their customers' needs, expectations, and demands, leading to smarter, modernized, efficient and reliable energy systems that deliver lower prices, fewer outages, and lower emissions to the customers. For example \cite{Chima2011}, the company OPower creates a demographic profile in Arlington, VA based on energy consumption data from its smart meters, and groups similar households into communities. OPower then allows these groups to compare their energy usage against each other and compete head-to-head to see who can reduce energy consumption the most. Social network for smart grid (SSG) would serve as a strong pillar in making smart grid smarter by including the customers and their real time peer-to-peer data sharing.

However, this also opens up a channel for the attackers to attack the power grid utility service. Attacks through social network can use the crowd-shifting strategy as well as information propagation to spread misinformation among the customers. This includes false electricity prices where the customers might increase their load given a lower price and shut off their utilities given a higher price. Misinformation such as future power failure or breakdown can also trick the customers to alter their power usage. As the integration of social network into smart grid is inevitable, so are the problems associated with the integration. Although social network misinformation attacks have been studied \cite{Wagner2012, Nguyen2012, Reid2016}, its effects were not extended to the critical infrastructure. Thus, considering the possible catastrophic impact on the smart grid when demands on loads are altered \cite{Mishra2015}, it is crucial to study and analyzing the vulnerability of smart grid from the coupled social network. 

In this paper, to better understand the vulnerability of smart grid, we aim at finding the most critical nodes in the coupled social network, such that when those nodes believe in the misinformation on smart grid, they may spread the misinformation to a large portion of nodes in the social network and in turn results in severe failure in the smart grid. We term the problem as Misinformation Attack Problem in Social-smart grid (MAPSS). The critical nodes found in MAPSS can guide the utility companies and the social network administrators to apply corresponding precautions to failures. 

Nonetheless, MAPSS is challenging as we need to consider not only the complicated information diffusion phenomenon and power network dynamics themselves, but also their interdependency. If we only consider maximizing the misinformation diffusion, the result can be suboptimal as the influenced nodes may not be critical in the smart grid. Also, it is hard to find critical nodes in the smart grid itself for MAPSS, as the constraint on the number of critical nodes is for the social network and we have no prior knowledge on number of critical nodes for the smart grid. To cope with the challenges, we propose techniques that can obtain the criticality of nodes and set of nodes and then calculate the most critical nodes in an integrated fashion.


Our contributions are summarized as follows:
\begin{itemize}
	\item We propose and define a new research problem MAPSS that identifies the most critical nodes in social network coupled smart grid, which helps in understanding the vulnerability of this new system.
	\item We propose various attack strategies to the challenging problem of MAPSS. The Greedy Social Attack (GSA) can be useful when little of the smart grid is known. When smart grid information is available, the Social Power Attack (SPA) can explicitly consider cascading failure and calculate the criticality of the subset of social nodes based on their impact to the smart grid. SPA has two variations, SPA-Concurrent (SPA-C) attacks all the nodes at the same time; when a longer attack period is allowed, SPA-Sequential (SPA-S) launches attacks in rounds, the decision of each round is based on current network status, and thus can result in severer failures.
	\item We introduce Controlled Load Shedding (CLS) as a protection strategy which stabilizes the system by load curtailment and load shedding.
	\item We test our attack strategies in various simulation settings with realistic social network and power network datasets and proved their effectiveness.
    \end{itemize}

The rest of the paper is organized as follows. We review the existing efforts on integrating the SNs with smart grid in Section \ref{sc:integration}. In Section \ref{model}, we discuss the models about the power grid, the information diffusion in the social network and the integrated social/power network. We also define MAPSS and prove its NP-hardness. The attack and protection strategies are described in Section \ref{sec:attack} and \ref{sec:protect} respectively. We present the experiments in Section \ref{eval} and conclude the paper in Section \ref{conc}.

\section{Social Network Integration}\label{sc:integration}

Smart grid promises to offer energy and money savings to both utility companies and consumers. Different pricing programs used by utility companies had performed well at small scale in opt-in trials, but had consistently failed to successfully scale \cite{Davito2010}. For example, Exelon companies Baltimore Gas and Electric (BGE) and Commonwealth Edison (ComEd) chose to deploy peak-time rebate (PTR) programs capable of reaching all of their customers. 
They both realized that price alone was insufficient to scale dynamic pricing programs. A close study of similar efforts in California showed that lack of customer awareness led to insignificant program participation and, in turn, program impact.

Thus, customer interaction and continuous real time engagement play an important role in achieving the goal set by smart grid along with the right technological advances. Social Network has already shown the world its power in the above requirements and is demonstrated in the literature as shown in Table \ref{tb:impact}.

\begin{table}[h] 
	
	\centering
	\caption{Impact of social network}
	\resizebox{9cm}{!} {
		\begin{tabular}{ | c| p{6cm} | }
			\hline
			Literature & Impact of SSG \\
			\hline
			\cite{Stein2012} & Smooth load curve for households, hence no extra costly power plant capacity will be required. \\
			\cite{Chat2013} & 6\% reduction in peak load, aggressive incentives reached a 14\% reduction in peak load. \\
			\cite{Lei2012} & Consumers reduced their annual energy usage by an average of 2.8\% when given comparison information with other people’s energy consumption.\\
			\cite{Skopik2014} & Energy savings of 7-9\% by mutual sharing through social network.\\
			\hline
		\end{tabular}
	}
	\label{tb:impact}
\end{table}
And rightly so, utility companies have started integrating social network to connect their customers. Some of the existing social networking tools available are
GreenPocket (sharing experiences and contest participation), OPower (Facebook based community approach to sharing energy saving tips), Ensemble (customer interaction through incentives and competitions).

Social network strives on the principle that \emph{people want to do what others like them are doing}. This forms the basis of the customer engagement and interaction and drives towards achieving energy efficiency and savings. However, using this crowd-shifting theory, attackers could propagate false information, engaging customers in time critical competitions all directed to change load profiles of customers in a community hidden from the utility companies. The attack can be significant enough, when combining with other such attacks, to produce line failures leading to massive blackouts. In our observation, this is the first work of its kind to consider the effect of social networking in the power grid and understand the real impact of misinformation propagation in the social network on the power grid.

Electricity consumers may be users in the social networking tools developed by the utility companies. As opposed to the normal information propagation, the users might be involved in sharing or forwarding message \cite{Huang2015} such as
(1) A general information, (2) A load-shifting tip, and (3) An energy-reduction tip. In this paper, we are concerned with the message types (2) and (3) and the impact of the misinformation associated with the above types of messages in the social network.

\begin{figure*}[t]
	\includegraphics[width=1\textwidth]{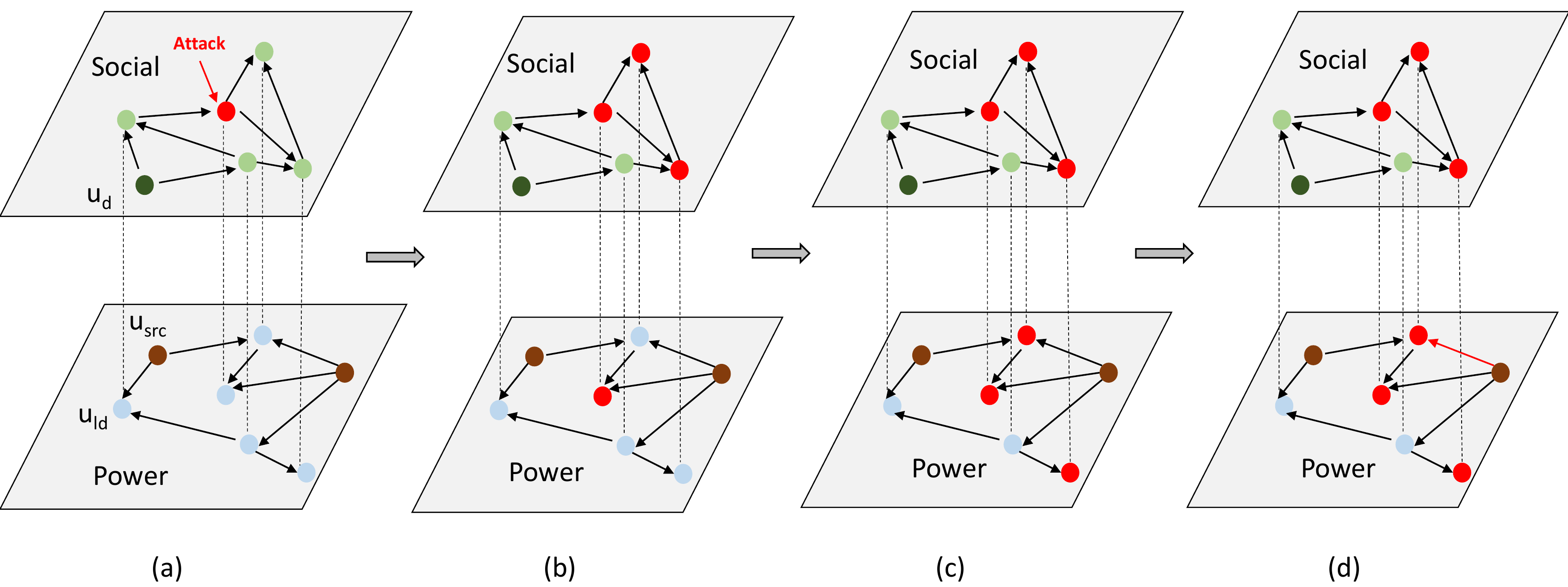}
	\centering
	\caption{In this figure, the model as well as the problem MAPSS are discussed. In Fig. (a), the social network $G_S$ with its users $V_S$ and the connections in between them $E_S$ are represented by the directed edges and each user $u_S$ has a corresponding node in the power network $G_P$. $u_d$ represents the utility company in the social network. Note that attacking the SN through $u_d$ is not feasible as the company can easily identify the attack, so we do not consider it. In the power network, nodes are represented by $u_{ld} \in V_{ld}$ and $ u_{src} \in V_{src}$. Note there could be power nodes without their social network profile. The attack starts with a user in a social network i.e. seed node. In Fig (b), the corresponding node in the power network starts increasing the load. Meanwhile the information starts spreading and influences other nodes in the social network. The corresponding failure in the power grid is represented by Fig (c). Because of the numerous load changes, it might to thermal and cascading failure leading to failure of other edges and nodes as shown in Fig (d).}
	\label{fig:model1}
\end{figure*}

\section{Model} \label{model}
In this section, we describe the models for information diffusion in the social network, model for the power grid network and the overall model connecting the users on the social network to their corresponding load nodes. {\color{blue}The notations used in this section are summarized in Table~\ref{table:notation}.}

\begin{table}[h]
\caption{Summary of Notations for Section \ref{model}}
\label{table:notation}
\resizebox{9cm}{!} {
\begin{tabular}{ |l|p{6cm}| }
	\hline
	Variable & Meaning \\
	\hline
	$G_S$ & $G_S=(V_S,E_S,w)$, the social network \\
	$G_P$ & $G_P=(V_P,E_P)$, Power grid network with nodes and transmission lines. \\
	$P$ & Set of power generation nodes.\\
	$D$ & Set of demand nodes.\\
	$p_i$ & Power generation output of node $i$.\\
	$d_i$ & Load demand of node $i$.\\
	$f_{ij}$ &Power flow in transmission line $(i,j)$.\\
	$u_{ij}$ & Capacity of transmission line $(i,j)$.\\
	\hline
\end{tabular}
}
\end{table}

\subsection {Social Network Model}

We model the social network as a weighted directed graph $G_S=(V_S,E_S,p)$ with a node set $V_S$ and a directed edge set $E_S$, where a node $v \in V_S$ represents a user and an edge $(u,v) \in E_S$ exists if and only if node $v$ is connected to node $u$, that either node $v$ can receive the information shared or directly sent from $u$. Also each $(u,v)$ is associated with a weight $p_{uv} \in [0,1]$ to denote the probability of information propagation, which is discussed below.

\paragraph*{Information Propagation}

In order to model the information propagation in the social network, we will focus on the Independent Cascading (IC) model which is widely adopted \cite{Kempe2003}. However, the results can be easily extended to other propagation models such as the Linear Threshold (LT) model. 

In the IC Model, initially no nodes adopt the misinformation. Given a seed set $S$, the misinformation diffusion proceeds in rounds. In round $0$, all nodes $v \in S$ are influenced by the misinformation and all other nodes remain uninfluenced. In round $t \geq 1$, all nodes influenced at round $t-1$ will try to influence their neighbors based on the edge weights. A node $u$ influenced at round $t-1$ has probability $p_{uv}$ to influence an uninfluenced neighbor $v$ at time $t$. $u$ cannot influence any neighbors at any time $t^\prime > t$ and it stays influenced till the end. The process stops when no more nodes can be influenced. We denote $I(S)$ as the expected number of nodes influenced by the misinformation with $S$ as the seed set, where the expectation is taken over all $p_{uv}$. We call $I(S)$ as the influence spread of seed set $S$.

\subsection {Power grid Model}\label{ssc:powergrid}

We use the linearized DC power flow model to describe the power flow in the power grid. In the linearized approximation, we are given a power grid represented by a directed graph $G_P=(V_P,E_P)$, where:
\begin{itemize}
	\item Each node $i \in V_P$ corresponds to either a power generator (i.e., a supply node), or to a load (i.e., a demand node), or to a node that neither generates nor consumes power.
	
	\item  If node $i$ is a generator and it is operating, then its output must be in the range [$P^{min}_i$, $P^{max}_i$] where $0\leq P^{min}_i\leq P^{max}_i$; if the generator is not operated, then its output is zero. The set of generators is denoted by $P$.
	
	\item If node $i$ is a demand, then the "nominal" demand is given by $D^{nom}_i$. The set of demand nodes is denoted by $D$.
	
	\item The edges $E_P$ represent power/transmission lines. For each line $(i, j)$, two parameters are given i.e. $\pi_{ij} > 0$ (the resistance or reactance) and $u_{ij}$ (the capacity).
\end{itemize}

Now, given a set $P$ of operating generators, the linearized power flow is a solution to the system of constraints given in the following set of the equations. For each edge $(i,j)$, $f_{ij}$ represents the power flow on the edge (transmission line) $(i,j)$. In the case where $f_{ij} < 0$, power is effectively flowing from $j$ to $i$. Additionally, the phase angle at node $i$ is given by the variable $\theta_i$. Given a node $i$, $\delta^+(i) (\delta^-(i))$ is the set of lines oriented out of (into) node $i$ and $N(i)$ is collection of nodes connected to $i$.
The power flow equations are given below:
\begin{align}
	\sum_{(i,j) \in \delta^{+}_{i}}f_{ij} - \sum_{(j,i) \in \delta^{-}_{i}}f_{ji} & = 
	\left\{ 
	\begin{array}{l l}
		p_i & \quad i \in P\\
		-d_i & \quad i \in D\\
		0 & \quad \text{otherwise}
	\end{array} 
	\right. \label{pf1}
\end{align} 
\begin{align}
	\theta_i - \theta_j -\pi_{ij}f_{ij}=0, &&\quad\forall (i,j)\in E_P\label{pf2}\\
	p_i^{min}  \leq  p_i \leq  p_i^{max}, &&\quad \forall i \in P \label{pf3}\\
	0 \leq  d_j \leq d_j^{nom}, &&\quad \forall j \in D \label{pf4} 
\end{align}

\subsubsection{Cascading failure model}

Next, we describe the cascading failure model in Algorithm \ref{alg:template}. This is an extension of the model in \cite{Mishra2015}. In the event of line failures, there might be formation of components, that are disconnected from each other. Following this, we adjust the total demand equal to the total supply. We then recalculate the power flows in the network using the power equations. The new flows may exceed the capacity and as a result, the corresponding lines will become overheated and thus vulnerable to failure. We assume that the cascade proceeds without encountering any intervention from the system operator for the majority of the paper, to evaluate the worst-case vulnerability of the smart grid. In order to model system intervention, we develop a protection strategy in parallel.
\begin{algorithm}[h]
	
	\caption{Cascade Failure Model}
	\label{alg:template}
	\KwData{Connected Power grid Network $G(V,E)$}
	\KwResult{$S_1$: Lines which failed, $S_2$: Nodes which failed}
	\While{Network is not stable}{
		Adjust the total supply to the total demand within each island.\\
		Use equations (1)-(4) to calculate power flows in G. \\
		For all lines, compute the moving average 
		$\tilde{f}^t_{ij} = \alpha f_{ij} + (1-\alpha)\tilde{f}^{t-1}_{ij} \makeatother$.\\
		Remove all lines that have moving average flows greater than the capacity ($\tilde{f}^t_{ij} > u_{ij}$) and add to $S_1$.\\
		Add the failed nodes to $S_2$.\\
		If no more line fails, then network is stable, break the loop.
	}
\end{algorithm} 

\subsubsection{Line failure} Similar to \cite{Mishra2015}, the transmission line failure is caused by overloaded lines. The outages are modeled based on moving average flow $\tilde{f}^t_{ij}$, where $\tilde{f}^t_{ij} = \alpha f_{ij} + (1-\alpha)\tilde{f}^{t-1}_{ij}$. A general outage rule gives the fault probability of the line $(i,j)$, given its moving average $\tilde{f}^t_{ij}$. In this paper, line failure happens when $\tilde{f}^t_{ij} > u_{ij}$.

\subsection{Interdependency Model}
In this section, we discuss the interdependency between the social network and power grid as shown in Fig. \ref{fig:model1}. In the smart grid, social network will be used as the tool to actively involve the consumers in through channels such as peer to peer communication, community based discussion and user experience forums. The customers can engage in demand and response programs and actively share energy data among themselves. This not only helps the utility distributor in energy management but also helps in making smart grid really "smart". This data can be shared through a dashboard that offers user-friendly data displays, such as graphs of energy consumption over time, which allow for remote monitoring of energy use from an office computer, tablet or smart phone. Information such as where energy is being spent, the benefits of conservation or efficiency measures, as well as anomalies, such as energy usage levels that fall outside of norms or expectations will be made available at the real time.

Every node $u_P \in D\subseteq V_P$
\footnote{Note that in addition to the load nodes being a part of the social network, the utility distribution company and power generation companies are also the members of the social network. By this they are able to send important information to the customers regarding the demand supply, price information and other real-time usage statistics of the power grid.} is a customer in the power utility service (such as a residential house, an industrial building, etc) and is associated with the social network account represented by user $u_S \in V_S$. We denote the collection of correspondence $(u_P,u_S)$ as $E_{PS}.$ Social networking gives the opportunity to users $u_S \in V_S$ to add other users and community members as friends in order to share their energy usage, offers, pricing etc. Utility companies can use the social network to denote change in price, real-time demand and supply and issue any warnings as required, in order to changing the demand pattern of the customers in the power network via crowd-shifting in the SN. Once a user accepts the information coming from its neighbors (i.e. is "influenced") s/he alters the load in the house in real-time to save money or equipments (in case of potential power cuts and surges).

The attackers might exploit the social network designed for keeping customer active in the smart grid system, via spreading misinformation about price changes, power surges etc. In this paper, we consider a specific type of attack (misinformation attack) that in the social network, the attacker might spread information about the lowering of the electricity price for a few hours during the day. This would lead to users increasing their load at their residence during those times to utilize the lower rates, which is followed by sudden spikes in the demand, making the system unstable. A coordinated attack as this would lead to line failures and thus cascading failures causing massive blackouts and equipment damages. 

Consider the power grid $G_P$ as discussed in Sect.~\ref{ssc:powergrid}. We shall use the notation $F(G_P)$ to denote the number of nodes in $D$ which do not have a path to any node in $P$ in $G_P$. That is, they are disconnected from the sources of power generation and hence considered as failed nodes. 

With the given models in mind, we define the MAPSS problem as follows:

\begin{mydef}[MAPSS]
Given the social network $G_S = (V_S,E_S,p)$, power network $G_P = (V_P,E_P)$ and edge set $E_{PS}$, identify $k$ nodes in $G_S$, whose activation would lead to maximum number of failed/disconnected nodes in $G_P$ based on misinformation attack.
\end{mydef}

MAPSS is NP-hard as its special case where $E_S=\emptyset$ is similar to the RAA problem defined in \cite{Mishra2015} and is proved to be NP-hard. Also, using the result from the same paper, MAPSS cannot be approximated within a factor of $O(|E_P|^{1-\eta})$ for any $\eta>0$, unless P=NP. 
 
\section{Attack Strategies} \label{sec:attack}
In this section, we propose several approaches for MAPSS. The first approach, Greedy Social Attack (GSA), greedily selects social nodes iteratively, maximizing the marginal gain of social node activation $I(S)$ and calculating the impact to $G_P$. This approach is useful when the knowledge on $G_P$ is limited to the attacker. When $G_P$ is available to the attack, two more advanced approaches are proposed to launch Social Power Attack (SPA) considering cascading failure during social node selection. The difference between the two methods is on whether the nodes are selected concurrently (SPA-C) or sequentially (SPA-S). 
\subsection{Greedy Social Attack (GSA)} 
The motivation behind this attack is to spread misinformation in the SN as much as possible without concerning about corresponding impact of load node failures in the power network. Denote $V_S^p$ as the set of SN users corresponding to power nodes in $G_P$. Then, we can utilize targeted influence maximization algorithms such as BCT \cite{Nguyen2016} to calculate the desired seed set. The failure in $G_P$ can be obtained by first retrieving the influenced nodes in $G^S$ and then applying Alg.~\ref{alg:template} in $G_P$. {\color{blue}We describe the details of GSA in Alg.~\ref{alg:SG}.}
 
\begin{algorithm}[h]
	\caption{Greedy Social Attack (GSA)}
	\label{alg:SG}
	\KwData{$G_S(V_S,E_S), V_S^p\subseteq V_S$,k}
	\KwResult{$S, F(G_P)$}
	Initialize $S = \emptyset$\\
     Calculate $S, |S|\leq k$ based on the algorithm BCT \cite{Nguyen2016} with uniform cost and $V_S^p$ as the target set. Set benefit for all $v\in V_S^p$ as $1$ and benefit for all other nodes as $0$.\\
	\textbf{Return} $S$
\end{algorithm}

GSA can be launched when the attacker only gains access to the social network. Hence, the advantage of using this algorithm is the attacker can launch the attack without the knowledge of the load profiles, the power network topology and other statistics which might have been hidden from general users of the social network deployed by the utility companies. 

\begin{algorithm}[h]
	\caption{Cascading Impact Calculator (CIC)}
	\label{alg:CIC}
	\KwData{$G_S(V_S,E_S), G_P(V_P,E_P), S$}
	\KwResult{$CI$}
    \For{$i\in V_P$}
    {
      Calculate $ci_i.pload$ \Comment{Nodes to attack in $G_P$ to fail $i$}\\
      Calculate $ci_i.nodes$ by Alg \ref{alg:template} \Comment{Damage when $ci_i.pload$ nodes are attacked.}\\
      $A_S = ci_i.pload \rightarrow V_S$ \Comment{Project power to social}\\
      \While{$|A_S| < |I_{A_S}(S^\prime)|$}{
          select $u_S = argmax_{v_S \in V_S\backslash S^\prime}(I_{A_S}(S^\prime \cup {v_S}) - I_{A_S}(S^\prime))$\\
          $S^\prime = S^\prime \cup \{u_S\}$\\
      }
      $ci_i.seeds = S^\prime - S$\\
    }
	\textbf{Return} $CI$
\end{algorithm}
\subsection{Social Power Attack (SPA)}
Although GSA can be launched quickly, the lack of knowledge of its impact on the power grid does not result in large cascading failures, rather just high misinformation spread. In this section, we discuss SPA which explicitly takes cascading failure into consideration and focusing on to maximize the number of node failures in the power network. It has two variations, SPA-Concurrent (SPA-C) tries to select seed nodes concurrently and SPA-Sequential (SPA-S) tries to identify seed nodes iteratively after assessing system state due to previous failures. To evaluate cascading failure resulted from node failures, we first propose the Cascading Impact Calculator (CIC) in Sect.~\ref{ssc:cic} and then discuss SPA-C and SPA-S in subsequent sections.

\subsubsection{Cascading Impact Calculator}\label{ssc:cic}

In order to maximize the cascading impact of the attack, the Cascading Impact Calculator (CIC),  as described in Alg.~\ref{alg:CIC}, calculates the cascading impact of all nodes upon its failure, given the current status of two networks and the selected seed nodes $S$. CIC returns the cascading impact list $CI$ that contains the cascade impact $ci_i$ for each node $i\in V_P$, including the seeds $ci_i.seeds\subseteq V_S\backslash S$ required to fail $i$ and the set of failed nodes $ci_i.nodes\subseteq D$ when $i$ fails. Function $I_{A_S}(S^\prime)$ calculates the number of influenced nodes in $A_S$, which is the projected set of nodes from the power network. $S^\prime$ holds the seed nodes that are required to fail $ci_i.nodes$. 
\begin{figure}[!b]
\begin{align}
	 \min & \sum\limits_{i \in D}z_i& \label{ip1} \\
	\text{s.t. }
    &t_o = 1\label{ip2}\\
	& \text{Equations } \eqref{pf1}-\eqref{pf4}\label{ip3}\\
    &d_i = d^0_i (1 + z_i*\Delta_i),\quad  \forall i \in D& \label{ip7}\\
    &p_i = p^0_i + \Delta p, \quad \forall i \in P&\label{ip:generator}\\
	&f_{ij} + M \omega_{ij} \geq u_{ij}y_{ij},\quad  \forall ij \in E_P& \label{ip51}\\
    &-f_{ij} + M (1-\omega_{ij}) \geq u_{ij}y_{ij},\quad  \forall ij \in E_P& \label{ip52}\\
    &M(1-t_i) \geq \sum_{j\in N(i)\backslash P}(1-\phi_{ij}) \nonumber\\
    &+\sum_{j\in N(i)\cap P}(1-y_{ij}) ,\quad\forall i \in V_P\backslash P& \label{ip:connection}\\
    &\phi_{ij}\geq t_j ,\quad\forall j\in V_P\backslash P, i\in N(j)\backslash P&\label{ip:aux1}\\
    &\phi_{ij}\geq y_{ij} ,\quad\forall j\in V_P\backslash P, i\in N(j)\backslash P&\label{ip:aux2}\\
     &\phi_{ij}\leq t_j +y_{ij} ,\quad\forall j\in V_P\backslash P, i\in N(j)\backslash P&\label{ip:aux3}\\
	&z_i,t_i \in \{0,1\} ,\quad \forall i \in D& \label{ip9}\\
	&y_{ij},\omega_{ij} \in \{0,1\},\quad  \forall ij \in E_P&\label{ip10}\\
    &\phi_{ij} \in \{0,1\}, \quad \forall ij\in E_P || ji\in E_P&
\end{align}
\caption{The ILP for finding minimum number of nodes to attack in order to fail node $o$.}\label{fig:ilp}
\end{figure}

For calculating $ci_i.pload$, we construct the Integer Linear Program (ILP) as in  Fig.~\ref{fig:ilp} for failing the target node, denoted as $o$ in the ILP. The objective (\ref{ip1}) asks for the minimum number of nodes to attack so that node $o$ is failed. The binary variable $z_i$ reaches 1 if the node $i$ is attacked. Constraint (\ref{ip2}) guarantees the failure of node 
$o$ ($t_o=1$). Binary variable $t_i=1$ denotes that node $i$ is not connected to any generator. The power flow equations are given by constraint (\ref{ip3}). The demand of node gets directly affected by the attack, i.e demand rises in case of misinformation attack as shown in constraint \eqref{ip7}, where $d_i^0$ denotes the initial demand and $\Delta_i$ is the percentage of demand change when node $i$ is affected. The corresponding change in power generation is reflected in constraint \eqref{ip:generator}. When necessary, we increase the power generated by each generator by the same amount $\Delta p$ on top of the initial power generation $p^0_i$. Constraint (\ref{ip51}), \eqref{ip52} describes the case of power overflow, in which $M$ is a large constant, which can be set as the largest possible flow on a single line. The two constraints guarantee that if the absolute value of the flow $|f_{ij}|$ is beyond the line capacity $u_{ij}$, the line $(i,j)$ will fail (binary variable $y_{ij}=1$). The connectivity of demand nodes to sources are calculated in constraint \eqref{ip:connection} where the binary variable $\phi_{ij}=1$ denotes no connection for $i$ to $P$ through $j$, and is guaranteed by the auxiliary constraints \eqref{ip:aux1}-\eqref{ip:aux3}.

\subsubsection{SPA-Concurrent}

SPA-C uses CIC for the following:
\begin{itemize}
	\item $ci_i.nodes$ : Nodes in the power network that fail due to the failure of nodes in $ci_i.pload$. Let the sets of the failed nodes be represented by $\mathcal{P} = \{P_1,P_2,...,P_m\}$, where $P_i = ci_i.nodes$.
	\item $ci_i.seeds$ : Seed nodes in the social network required to spread information to targeted users in SSG and thus causing load imbalances in the power system leading to failure of $i$. We represent the sets of the seed nodes by $\mathcal{S} = \{S_1,S_2,...,S_m\}$ with $S_i = ci_i.seeds$.
\end{itemize}

\begin{figure}[t]
 	
 	\includegraphics[width=0.45\textwidth]{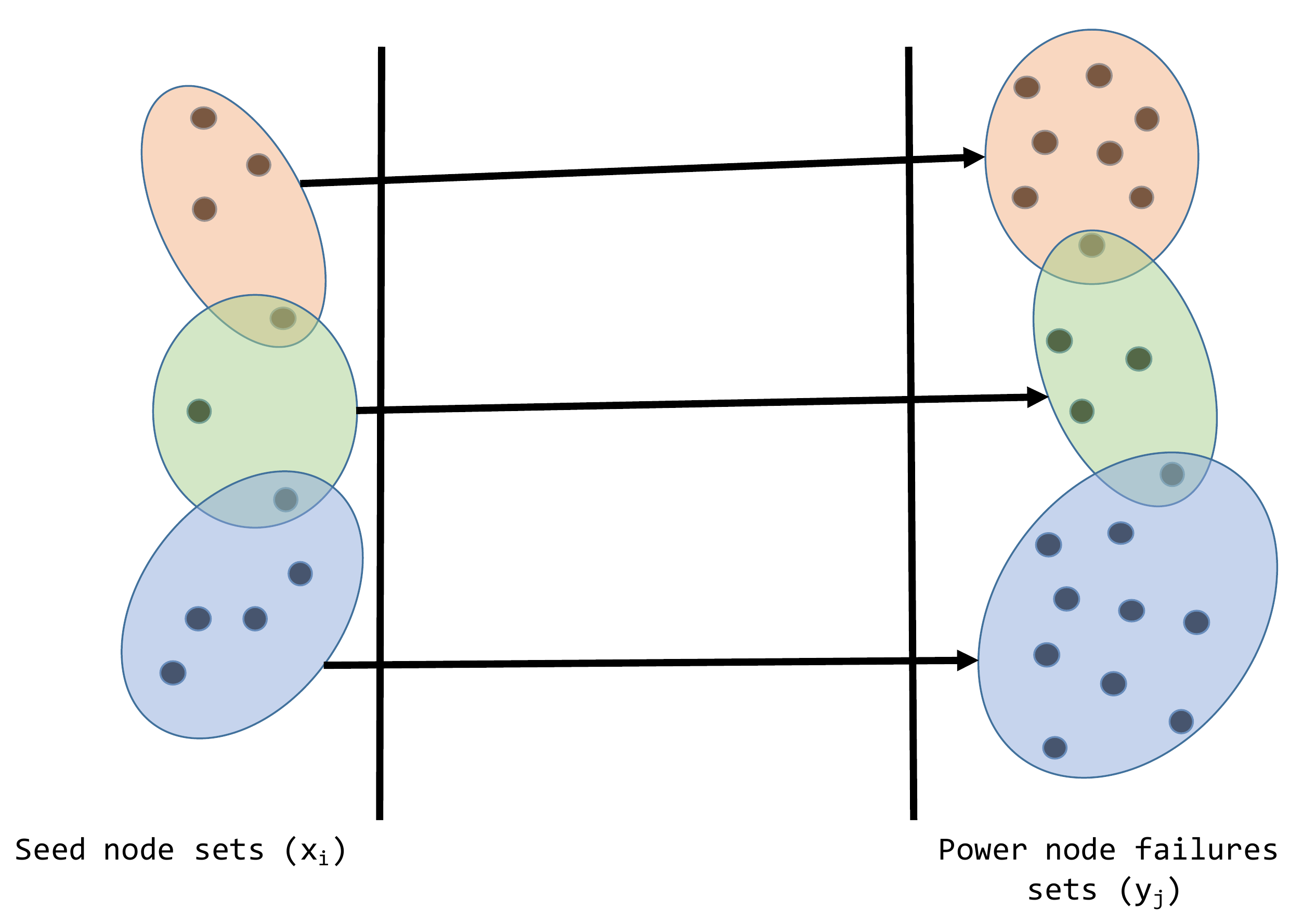}
 	\centering
 	\caption{SPA-C. The edges represent the correspondence of seed node failure with power node failures as given by CIC.}
 	\label{fig:spac}
 \end{figure}

Now SPA can be viewed as an arbitrary bipartite graph, with $\mathcal{S}$ social seed node sets on the left side, $\mathcal{P}$ power failure node sets on the right side and edges representing relationship between seed node selection and corresponding failure. The number of nodes in the power network is $m$, which determines the number of sets in $\mathcal{P}$ and $\mathcal{S}$. The seed node sets on the left side might overlap. Objective hence becomes to find minimum cardinality of seed node sets (left-side) which maximizes the covering failure node sets (right-side) in Fig.~\ref{fig:spac}. $x_i$ and $y_j$ are binary variables that represent the selection of node $i$ for seed set in social network and failure of node $j$ in power grid respectively. ILP formulation is given as below. In the formulation, constraint \eqref{eq:spa2} restricts the budget to be at most $k$. Constraints \eqref{eq:spa3} and \eqref{eq:spaadd} ensure the node failure status $y_j=1$ when any seed set $S_i$ that can fail $v_j$ ($v_j\in P_i$) is fully influenced. In constraint \eqref{eq:spa3}, $M$ is a large constant. 

\begin{align}
&\text{maximize } \sum\limits_{j:v_j \in V_P}y_j \label{eq:spa1} \\
&\text{s.t. }\sum\limits_{i: u_i \in V_S} x_i \leq k   \label{eq:spa2} \\
&M(2-y_j-\psi_{lj})\geq |S_l|-\sum\limits_{i \in S_l}x_i,\nonumber\\
&\qquad\forall j:v_j\in V_P, \forall l: v_j\in P_l\label{eq:spa3}\\ 
&\sum_{l:v_j\in P_l}\psi_{lj} = 1 \quad \forall j:v_j\in V_P\label{eq:spaadd}\\
&x_i \in \{0,1\}  \quad \forall i\in V_S\label{eq:spa4}\\
&y_i \in \{0,1\}  \quad\forall j\in V_P\label{eq:spa5}\\
&\psi_{lj} \in \{0,1\}  \quad\forall j:v_j\in V_P, \forall l: v_j\in P_l
\end{align}

Although SPA-C identifies the seeds optimally, it does not take into account system state changes which can alter the cascading impact of nodes. Since CIC calculates $ci_i$ independently of the failure of other nodes, it may lead to selection of less optimal seed nodes. Considering this, we propose the SPA-S.

\subsubsection{SPA-Sequential}

Unlike SPA-C, in which all the seed nodes are selected concurrently, the approach SPA-S involves iteratively: it chooses the seed nodes after assessing the failure caused by earlier seeds. CIC is called to re-evaluate the importance of each node in the updated and balanced power network following the seed node selection and the cascading failure. With that information, SPA-S (Alg.~\ref{alg:SPA2}) tries to greedily select the node with the highest $ci.nodes$ value, given the budget $k$ for the seed nodes is not exceeded. After each successful selection, $k^\prime$, $S$ and $G_P^\prime$ are updated which represent the budget, seed set and updated fault graph respectively. 

\begin{algorithm}[h]
	\caption{Social Power Attack Sequential (SPA-S)}
	\label{alg:SPA2}
	\KwData{$G_S(V_S,E_S), G_P(V_P,E_P), k$}
	\KwResult{$S$}
    $k^\prime = 0, S=\emptyset$\\
	\While{$k^\prime < k$}{
		$CI = CIC(G_P^\prime, G_S, S)$\\
		Sort $CI$ based on $ci.nodes$\\
		\ForEach{$ci_i \in CI$}{
			\If{$\#ci_i.seed < k-k^\prime$}{
				$S = S \cup ci_i.seed$\\
                $k^\prime = |S|$\\
				$G_P^\prime = G_P^\prime - ci_i.nodes$ \Comment{Remove failed nodes}\\ 
				break\\
			}
		}
	}
	\textbf{Return} $S$
\end{algorithm}

When there is no possibility to fail any node, we just add nodes from the social network to the seed set $S$ that minimize the yield of the power network as much as possible. This does not follow through with the goal of maximizing the load failure, but decreases the yield of the power network, which is a measure to understand the impact of attack on the power grid.

\section{Protection Strategy} \label{sec:protect}

This section describes protection strategy from cascading failures due to the social network attacks. Prevention of such attacks in social network can be achieved through monitoring of the social network by the utility companies. However, there is always a probability of misinformation propagation leading to cascading failure. Also, optimal mitigation of cascading failure is impractical given the dynamics of the network, large size of practical power networks, social network influence and multistage model as well as other reasons \cite{Bienstock2011}. The problem becomes more challenging as many new renewable sources of energy come to the aid of power generation which leads to more dynamic generation. 

Controlled Load Shedding (CLS): Restoring the grid to stable state once the cascading starts is a challenging problem given the parameters as described above. However, by controlled load shedding, smart grid should be able to minimize the damage. The above can be either achieved by load curtailment (large industrial customers have agreement with utility companies, so that they can be instructed to reduce demand in order to balance the system) or by load shedding in case load curtailment does not stabilize the system. In order to keep the yield of the system as high possible, we introduce the following linear program for controlled load shedding.

\begin{align}
\text{minimize } & \sum\limits_{i \in D}\tau_i& \label{sip1} \\
\text{s.t. }
Equ & (1)-(4)  \label{sip3}\\
& D_i = D_i^t -\tau_i, & \forall i \in D \label{sip8}\\
& p_i = p_i^t  - \beta_i & \forall i \in P \label{sip6}\\
&f_{ij} <= u_{ij} & \forall (i,j) \in E_P \label{sip4}\\
&f_{ij} >= -u_{ij} & \forall (i,j) \in E_P \label{sip41}\\
& 0 < \tau_i \leq D_i^t & \forall i \in D \label{sip2} \\
& 0 < \beta_i  \leq  p_i^t &  \forall i \in P \label{sip5}
\end{align}
In the formulation, the superscript $t$ denotes the values at time $t$, as the damage controls depends when CLS is implemented. $\tau_i$ denotes the value of load shedding or curtailment at a demand node $i$ with the largest change in load bounded by $\Delta D_i^{max}$ in constraint \eqref{sip2} and the actual load change in \eqref{sip8}. Constraint \eqref{sip3} is the power flow equation. Constraints \eqref{sip5} and \eqref{sip6} represent the generator ramp down in case of the fault by the utility company. No further line failures are guaranteed by constraints \eqref{sip4}, \eqref{sip41}.

\section{Experimental Evaluation} \label{eval}
In this section, we evaluate how the attack through social network can impact the power system in various settings. We first discuss the experiment setup and data sets in the section \ref{eval:setup} and then demonstrate the impact of MAPSS and the protection strategy in the following sections.

\subsection{Setup} \label{eval:setup}
 Our goal in the experiments is to demonstrate the effectiveness and impact of MAPSS in social network enabled power grid. The real-world data sets are limited due to very recent emergence of this concept. However, we do have access to the data sets of individual systems (social and power) and other resources that have just started researching the social power network \cite{Huang2015,Cassidy2014,Anda2014}. So in order to represent our model, we choose power systems as the base layer and build the social network on the top of the power network. Power system datasets are chosen from MATPOWER library \cite{Zimmerman2011}. The datasets used for our experiments are Pegase 1354 bus test case, IEEE 30 bus test case and IEEE 300 bus test case.
 
 To model information propagation in the OSNs, we use real-world Facebook network topology from Stanford repository \cite{Leskovec2012}. We sample a random set of connected users (same as the number of users in the power network) from the Facebook dataset of 4039 users. The influence between a pair of users was assigned randomly from a uniform distribution. The reason for such a setup is that the social network for smart grid deployed for the consumers only rather than the full set of consumers. 
 
For the link between the power grid and the SN, we construct an edge $e(i,u) \in E_{PS}$ between each power utility consumer $i \in V_P$ and a SSG user $u \in V_S$. With each unlinked demand node $i$ in $V_P$, we randomly select an unlinked node $u$ in $V_S$ and then map them together. With each neighbor of $i$, we randomly pick a neighbor of $u$, check whether they could be link together, if yes then recursively do this step with neighbor of $i$ and $u$. Finally, each demand node in $V_P$ would be linked to one node in $V_S$.

For comparison, we consider our algorithms GSA, SPA-C, SPA-S, as well as a random algorithm that selects power grid users randomly and then applying targeted influence maximization algorithm in the SN to influence those users. 

\subsection{Vulnerability of smart grid} \label{eval:sna}

In this section, we demonstrate the existence as well as the damage caused by MAPSS on the power grid. The measure we considered in this section is the number of failed nodes, which is defined as the number of users which do not have a path through the active transmission lines connecting them to any of the power generation sources. 

\begin{figure*}[ht]
	\begin{subfigure}{.33\textwidth}
	  \centering
	  \includegraphics[width=.8\linewidth]{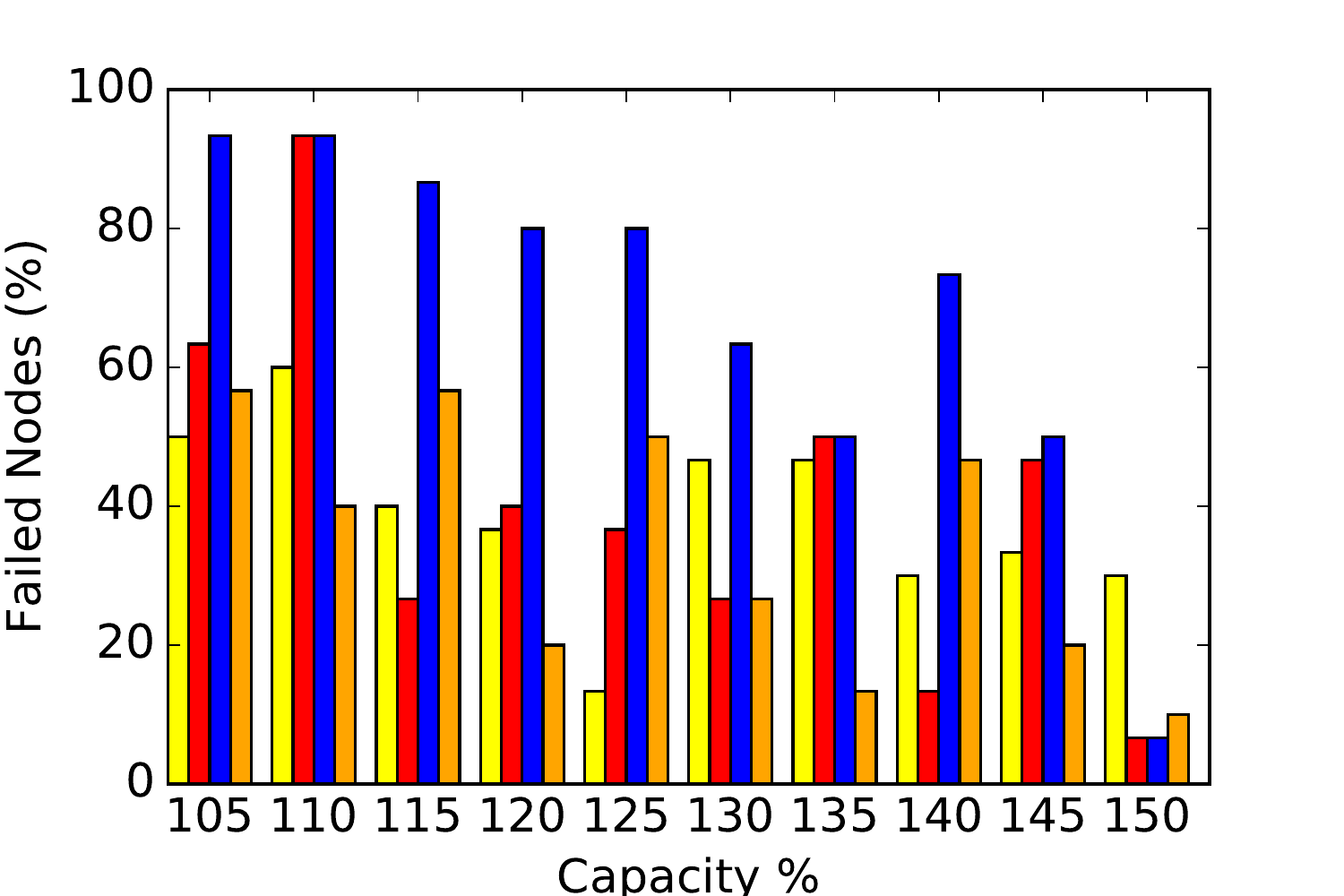}
	  \caption{IEEE 30 Bus}
 	  \label{fig:capacity30}
	\end{subfigure}
	\begin{subfigure}{.33\textwidth}
	  \centering
	  \includegraphics[width=.8\linewidth]{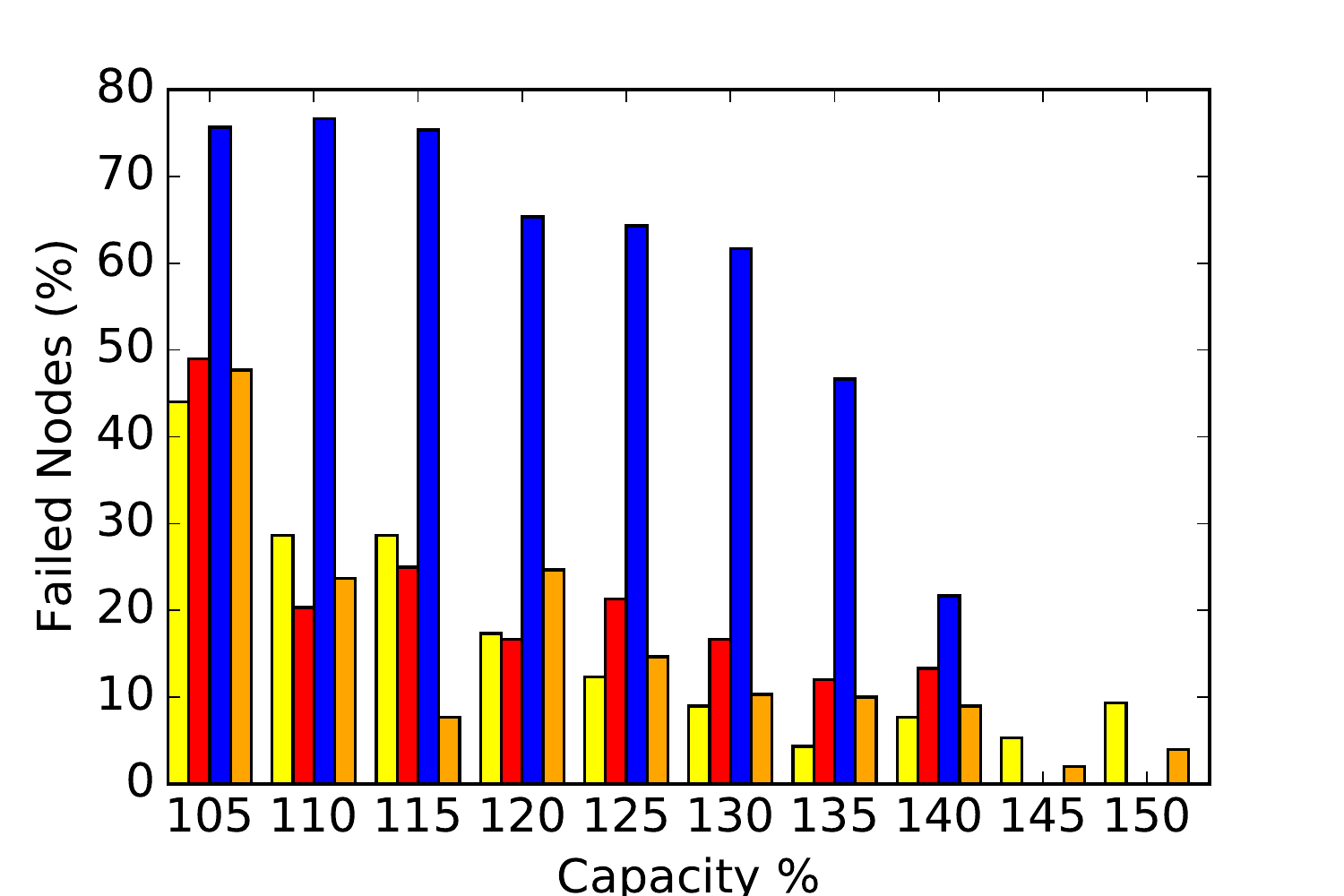}
 	\label{fig:capacity300}  
	  \caption{IEEE 300 Bus}
	\end{subfigure}
	\begin{subfigure}{.33\textwidth}
		\centering
		\includegraphics[width=.8\linewidth]{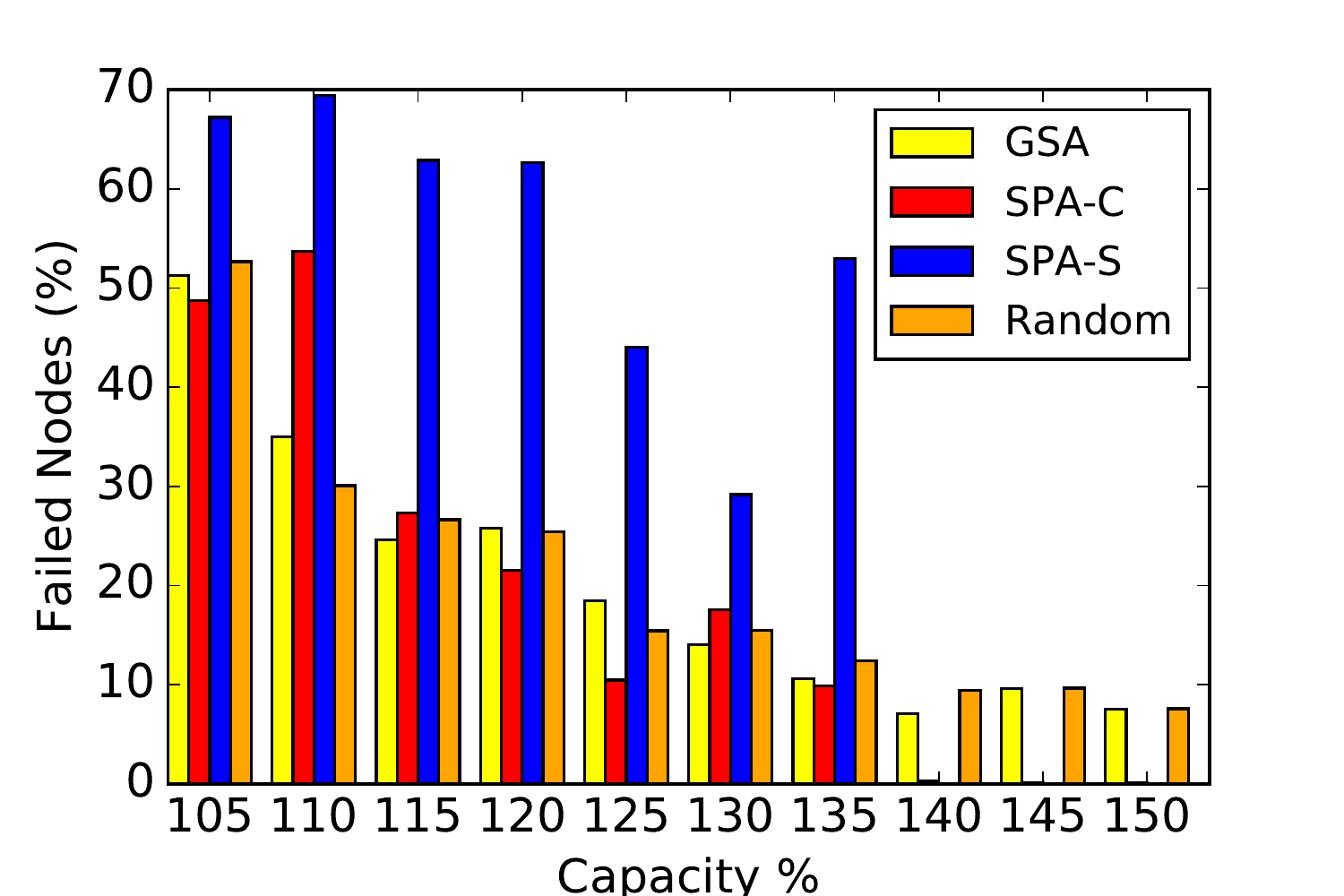}
 	\label{fig:capacity1354}  
		\caption{Pegase 1354 Bus}
	  \end{subfigure}
	\caption{Varying Line Capacity}
	\label{fig:fig}
\end{figure*}

In Fig.~\ref{fig:capacity}, we vary the capacity on the transmission lines. A $110\%$ capacity denotes that the capacity on each line is $110\%$ of the initial flow given by the dataset. We also fix the maximum possible load change at each user to be $25\%$ of its initial demand, according to the selection in \cite{Mishra2015}. The number of seeds is 5/20/50 for IEEE 30 bus/IEEE 300 bus/Pegase 1354 bus test cases respectively. From the result, we observe that all the algorithms are able to cause a large impact to the power network and the impact is subdued with higher capacity. Among the algorithms, SPA-S is capable of causing more severe damage compared with the others. For the other three algorithms, however, although SPA-C displays some superiority against GSA and Random, the performance of them are not stable. We also notice that the impacts of SPA-C and SPA-S drop to $0$ with high capacity on lines, which is due to the fact that the result of Alg.~\ref{alg:CIC}, $ci_i.seeds$ is too large.

In Fig.~\ref{fig:seed}, we fix the capacity at $130\%$, max load change at $25\%$ and varying the number of seed nodes allowed. Although SPS-S still constantly performs much better, the result may seem controversial as the percentage of failed nodes is not increasing with more seeds. However, it can be explained. One possibility is that as failure of transmission lines is mainly due to non-proportionally distributed demand, such scenario can happen when only a few users increased their demand. When more users increased their demand, the flow can even be more balanced. Thus, having more seeds may not guarantee a more severe failure. Another possibility is that in the social network, a very limited set of seed nodes is enough for influencing the majority of nodes, so that increasing the number of seeds will not have much impact.

\begin{figure*}[ht]
	\begin{subfigure}{.33\textwidth}
	  \centering
	  \includegraphics[width=.8\linewidth]{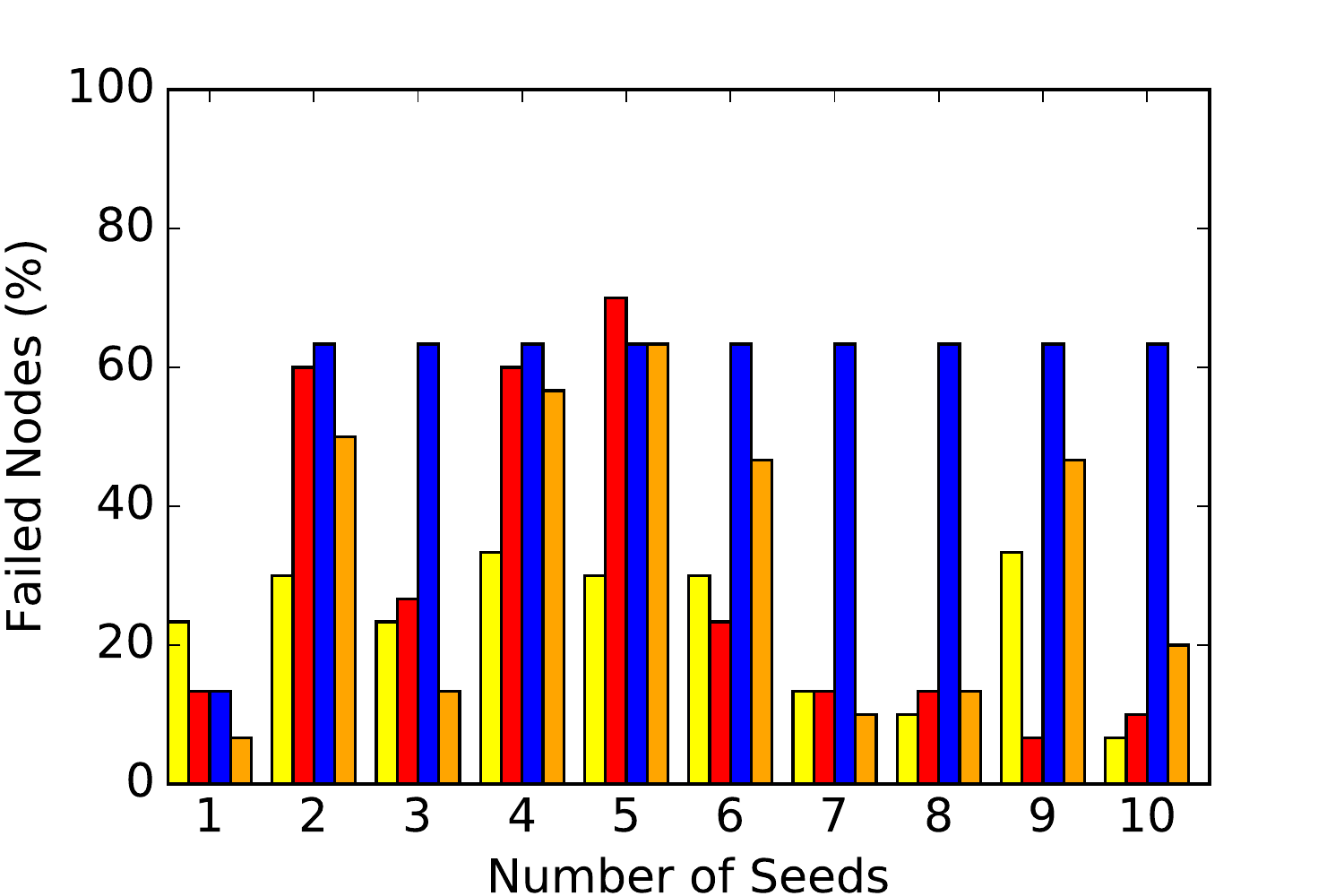}
	  \caption{IEEE 30 Bus}
 	  \label{fig:seed30}
	\end{subfigure}
	\begin{subfigure}{.33\textwidth}
	  \centering
	  \includegraphics[width=.8\linewidth]{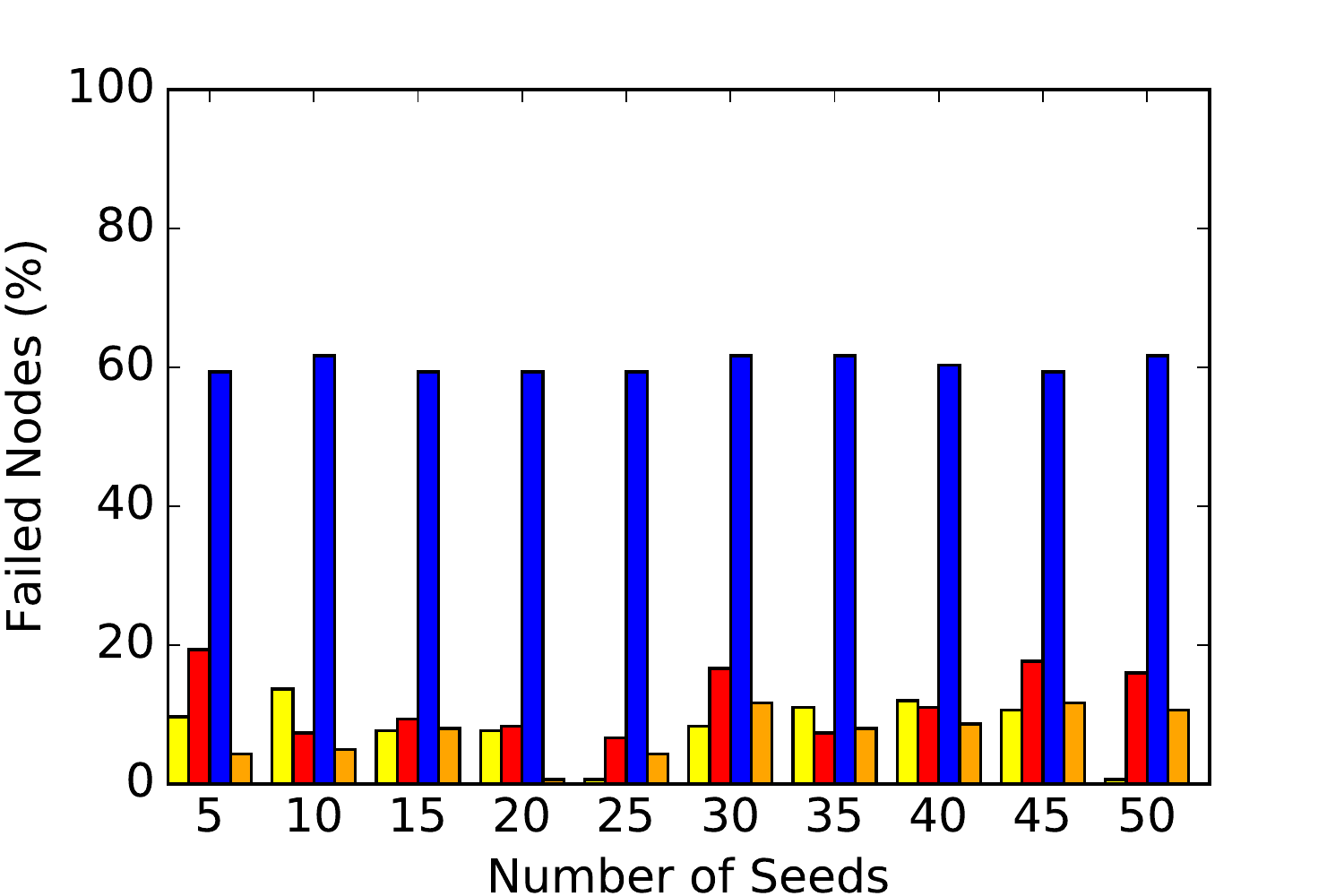}
 	\label{fig:seed300}  
	  \caption{IEEE 300 Bus}
	\end{subfigure}
	\begin{subfigure}{.33\textwidth}
		\centering
		\includegraphics[width=.8\linewidth]{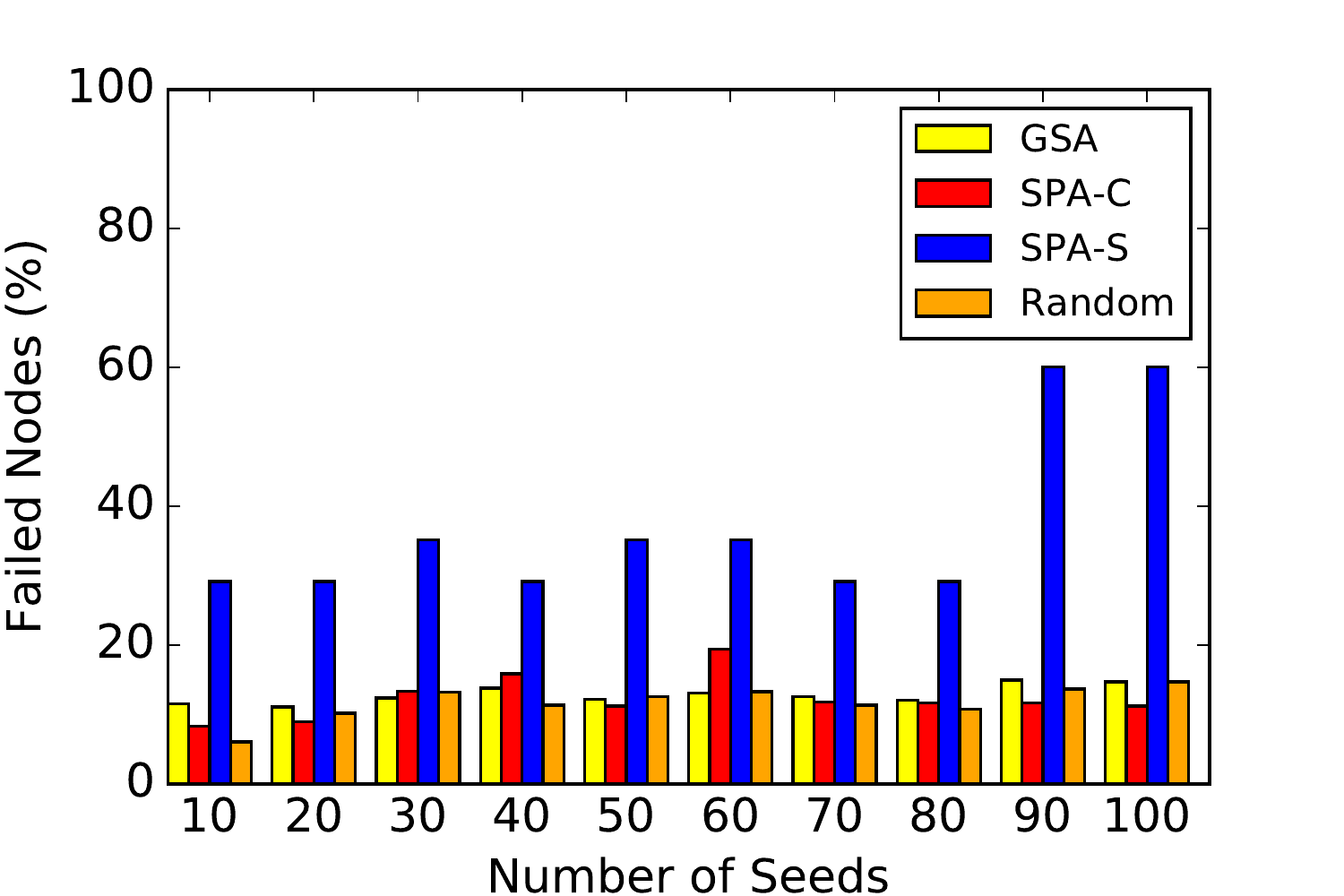}
 	\label{fig:seed1354}  
		\caption{Pegase 1354 Bus}
	  \end{subfigure}
	\caption{Varying Number of Seeds}
	\label{fig:fig}
\end{figure*}


\subsection{Controlled Load Shedding} \label{eval:CLS}

\begin{center}
	\begin{figure}[ht]
		\includegraphics[width=0.45\textwidth]{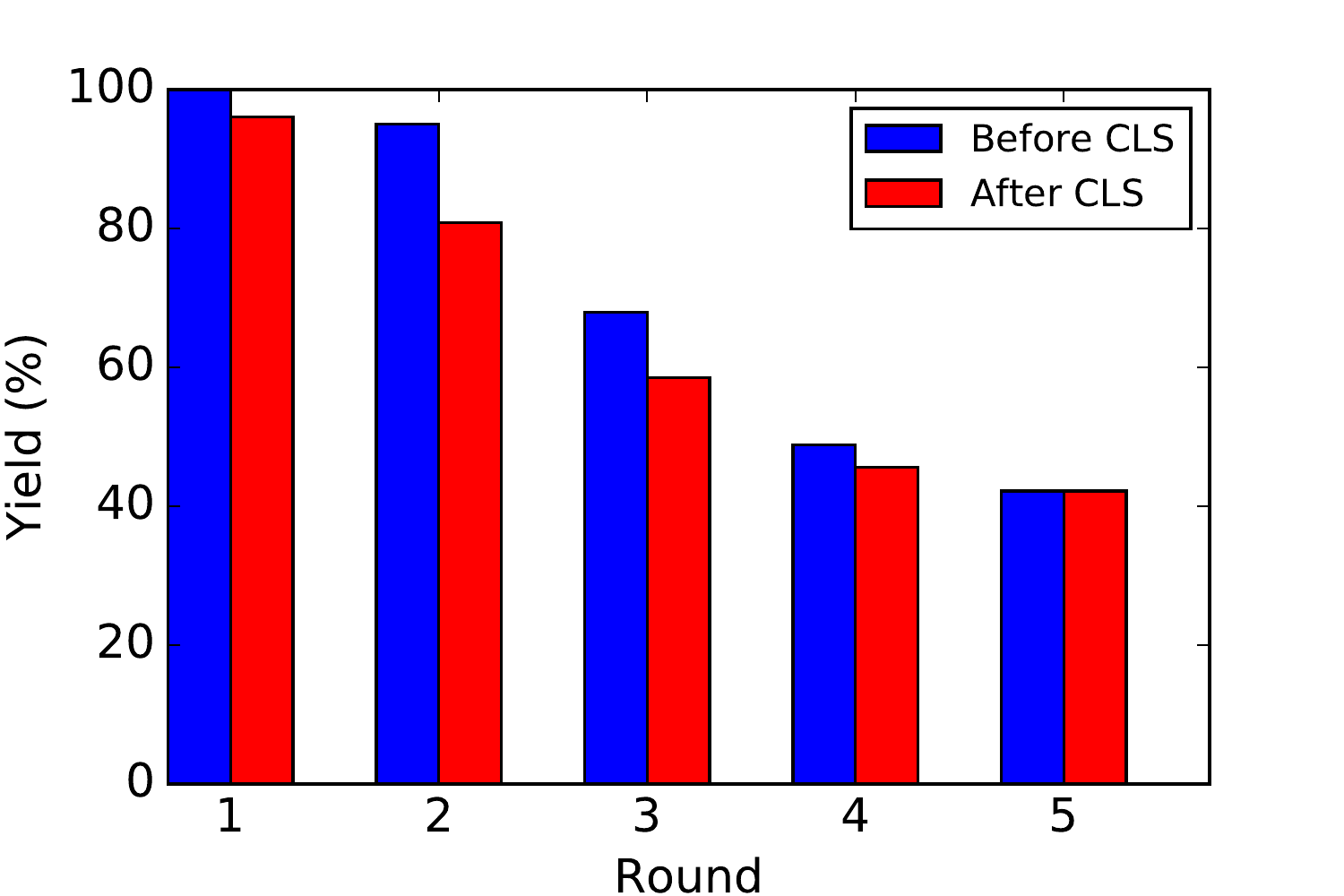}
		\caption{Protection by CLS. The yield of round 5 means no CLS is applied.}
		\label{fig:cls}
	\end{figure}
\end{center}

In this section, we evaluate the effectiveness of CLS and demonstrate that it depends on the state of the power grid when it is applied. In Fig.~\ref{fig:cls}, we consider a scenario in the IEEE 300 bus dataset when 10 users increased their demand and a five-round cascade was triggered. Before each round of cascade, we solve the CLS problem to obtain the minimum yield drop so as to prevent any further cascade. The yield is defined as:
	$\text{Yield} = \frac{\text{The actual demand at stability}}{\text{The original demand}}$. 
We use yield as the measure since load shedding will decrease the demand, which cannot be reflected by number of failed nodes. The yield naturally gives an assessment of the severity of the cascade as well as the cost to prevent further cascade. Clearly, CLS is more effective when applied to earlier rounds of cascade. Also, CLS can always help reducing the yield drop as long as the network is not stable.

\section{Conclusion} \label{conc}

In this paper, we analyzed the importance of social network in achieving the vision of smart grid. However, that comes with a price. We investigated new threats to the smart grid when social network is linked with it, by study the problem MAPSS and analyze the impact on the power grid. We also proposed the SPA algorithm which considers the cascading failure of the power grid in case of MAPSS. Misinformation propagation and cascading failure contribute to the severeness of these attacks as demonstrated by the experiments. For the future work, we will consider more user parameters for social network for smart grid and impact of failure of a substation on the social network connectivity. We are also interested in the protection strategies against MAPSS and smart attackers.

\section{Acknowledgements}\label{sc:acknowledgements}
This work was supported in part by NSF EFRI 1441231 and the Soonchunhyang University Research Fund.

bibliographystyle{unsrt}

\end{document}